# Photonuclear Reaction Cross Sections for Gallium Isotopes

Serkan Akkoyun[1], Tuncay Bayram[2]

[1]Cumhuriyet University, Vocational School of Healt, Sivas, Turkey

[2]Sinop University, Department of Physics, Sinop, Turkey

**Abstract**

The photon induced reactions which are named as photonuclear reactions have a great importance in many field of nuclear, radiation physics and related fields. Since we have planned to perform photonuclear reaction on gallium target with bremmstrahlung photons from clinical linear accelerator in the future, the cross-sections of neutron (photo-neutron ($\gamma$,xn)) and proton (photo-proton ($\gamma$,xn)) productions after photon activation have been calculated by using TALYS 1.2 computer code in this study. The target nucleus has been considered gallium which has two stable isotopes, $^{69}$Ga and $^{71}$Ga. According to the results, we have seen that the calculations are in harmony in the limited literature values. Furthermore, the pre-equilibrium and compound process contributions to the total cross-section have been investigated.

## Introduction

Photonuclear reactions have been used in basic and applied sciences in nuclear and radiation physics related fields [1-4]. When a gamma-ray is incident upon a nucleus, the excited nucleus behaves like any compound nucleus with an excitation energy. The most probable decay

*Corresponding author email address: sakkoyun@cumhuriyet.edu.tr*

*NUBA Conference Series-1: Nuclear Physics and Applications*

is neutron emission (γ,n). After this, (γ,2n), (γ,1p) and (γ,2p) reactions take part. Nuclear level and half-life identifications, nucleon binding energy determinations, material analysis, radiation protection applications, dosimetry, absorbed dose assessment, activation analysis, radiation transport analyses, physics of fission and fusion reactors, nuclear waste transmutations, and understanding element creations by astrophysical processes can be given as examples to such studies [2,4]. The experimental studies on these reactions have begun in 1934 [5] but there are still lack of existing data. Therefore, systematic studies of these reactions on different nuclei have been needed. The advantages of the reaction by photon activation are determination of the multiple element simultaneously, non-destructive structure of the process, requiring no time-consuming chemical separation procedure, deeper penetrating capability of the photon into the target [6]. The excited nucleus in the target emits particular radiation or photons in order to get rid of its excess energy. Due to the fact that photon do purely electromagnetically interaction with nuclei, the process is non-destructive.

Two types can be observed when two nuclear system collide and both are very important in order to understand nuclear reaction phenomena. In the compound nucleus reaction, two system collide and a highly excited intermediate system has been created. The compound system has been formed by a sequence of collisions leading to increasingly complicated rearrangements of the target nucleus. The excitation energy has been shared by nucleons. After sharing by chance, the overall energy has been localized on one or more or a group of nucleons for it to escape from the compound system. If there is still sufficient energy, further particle emission may occur, otherwise beta or gamma decay will appear. Second type which takes place very quickly according to the compound nucleus reactions has been named as direct reaction. In this type, there is no intermediate compound system formed. In the compound nucleus formation, the projectile passes into the interior of the nucleus suffering a number of collisions. Therefore, we have expected that the direct reactions are localized in the nuclear surface (projectile interact over the surface of the nucleus) whereas compound nuclear reactions are localized in the interior of the nucleus. For increasing energy or nuclei for which the compound nucleus does not have time to reach thermodynamic equilibrium, direct or pre-equilibrium processes may become significant.

The cross section of the pure compound process is very small with respect to the measured spectra. Additionally, the measured angular distributions are not isotropic as expected isotropic in compound nucleus decays. So there will be another classification between direct and compound nucleus mechanism (Fig.1). The pre-equilibrium process falls into either category. In these reactions, a nucleon is emitted after a number of collisions have occurred. Pre-equilibrium process [7] contributes to reaction cross-section in the 10 and 200 MeV energy range. In order to describe the pre-equilibrium mechanism, the exciton model is widely used [8]. At any moment during the reaction, the nuclear state is characterized by the total energy and the total number of particles above the Fermi surface and holes below the Fermi surface. Particles and holes are referred to as excitons. The complete formalism of the exciton model has been included in the TALYS reaction code [9]. By using the TALYS, it is possible to obtain all the compound nucleus, pre-equilibrium, and direct reactions contributions to the nuclear reactions.

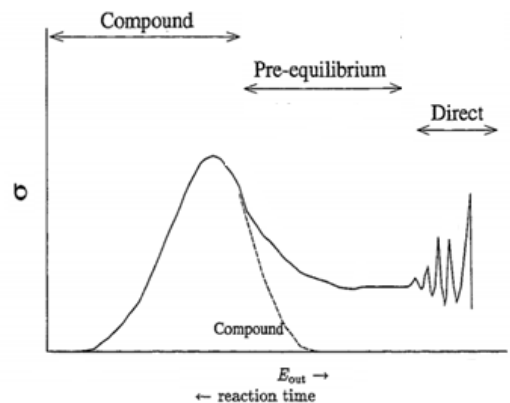

**Fig.1** Outgoing particle spectrum. The dashed curve distinguishes compound contribution from the other contributions. Outgoing particle energy increases to the right and reaction time increases to the left.

Gallium element has two stable isotopes and both are used in nuclear physics and nuclear medicine. $^{69}$Ga isotope has been used for $^{68}$Ge or $^{68}$Ga generators. The $^{68}$Ga which is a positron emitter has been used as a PET isotope. $^{71}$Ga has been used in the study of behavior of solar neutrinos and in NMR studies. In this work, the photo-neutron and photo-proton cross sections as a function of photon energy with one and two nucleon in out channels have been investigated by using the TALYS 1.2 computer program. The reactions considered were (γ,1n), (γ,2n), (γ,1p) and (γ,2p). According to the results, it has been seen that the calculations have been consistent with the TENDL 2013 values. There is very limited photonuclear reaction cross section data in the literature for gallium [10, 11].

In this work, the photo-neutron and photo-proton cross sections of gallium isotopes have been calculated according to the photon energy in the 0.5-30 MeV energy range by using TALYS 1.2 computer program. Default options have been used in the program. The results have been compared with the TENDL 2013 [12] and EXFOR [10] databases. Due to the lack of the experimental data in the literature for the photonuclear reaction cross section for the gallium isotopes, only (γ,xn) cross sections for natural gallium have been used for the comparisons. The compound nucleus and pre-equilibrium mechanism contributions have been also analyzed for gallium isotopes as done in a systematic study in reference [13].

**METHOD**

A linux based computer code TALYS has been used for the prediction and analysis of the nuclear reactions. TALYS code was written in Fortran computer language. Reactions involving neutrons, protons, gamma-rays, deuterons, tritons, helions and alpha particles have been simulated by TALYS program. The energy range for the incident particles has been between 1 keV and 200 MeV. In the code, the used nuclei mass for the target can be 5 and heavier. TALYS uses a suitable nuclear reaction model such as optical model, compound nucleus statistical theory, direct reactions and pre-equilibrium processes. In the code, the reaction cross section for all of the open channels can be calculated. Like nuclear level density, nuclear model parameter and gamma-ray strength function, several options can be included. In this study, default options have been used.

TALYS outputs include some information about the nuclear reaction such as elastic and inelastic scattering cross sections, total cross sections, angular distributions of elastic scattering, angular distributions in discrete levels, cross sections for isomeric and ground states, total particle energy and differential cross sections, emission cross sections, production cross section. In the output file, direct, pre-equilibrium and compound components of each cross section can be found.

**RESULTS and DISCUSSION**

In this work TALYS 1.2 version of the code has been used for the calculations of the (γ,1n), (γ,2n), (γ,1p) and (γ,2p) reaction cross-sections performed on both $^{69}$Ga and $^{71}$Ga targets separately. Pre-equilibrium and compound reactions cross sections have contributed to the overall cross sections. The energies of the gamma-rays have been 0.5 and 30 MeV with 0.5 MeV interval. The reaction Q values are -10.3 and -6.6 MeV for $^{69}$Ga(γ,1n)$^{68}$Ga and $^{69}$Ga(γ,1p)$^{68}$Zn reactions, respectively. Besides for $^{71}$Ga isotopes these value are -9.3 and -7.9 MeV for $^{71}$Ga(γ,1n)$^{70}$Ga and $^{71}$Ga(γ,1p)$^{70}$Zn reactions, respectively. Furthermore for two nucleon separations the Q values are -18.6, -16.6, -16.9 and -19.0 MeV for $^{69}$Ga(γ,2n)$^{67}$Ga, $^{69}$Ga(γ,2p)$^{67}$Zn, $^{71}$Ga(γ,2n)$^{69}$Ga and $^{71}$Ga(γ,2p)$^{69}$Zn reactions, respectively. These are the threshold energies for taking place the reactions. As can be seen in the figures that the cross sections are zero below these threshold values.

It is also clear in the Fig.2 that the cross sections for photo-proton reactions has been lower than that of the photo-neutrons because of the suppression by the Coulomb barrier. For instance, for 14 MeV incident photon energy on $^{69}$Ga target, the (γ,1n) reaction cross section is about 60 mb whilst the (γ,1p) reaction cross section is only about 3.3 mb for the same energy value. It is clear in Fig.2 that the photo-neutron cross section is almost 2 to 20 times larger than photo-proton cross section in the energy range 10.5 and 30 MeV. Since the threshold energy for the photo-proton reaction is lower than the photo-neutron reaction, there is a probability that occurring photo-proton reaction in the range of 7.5 and 10 MeV where photo-neutron cross is zero. As can be seen in the Fig.2 that after 19 MeV, (γ,2n) reaction become taking place. Until 20.5 MeV the (γ,1n) reaction cross section is larger than that of (γ,2n) reaction. But after this energy value, (γ,2n) become about 1.2 to 5.5 magnitude larger up to 30 MeV. Furthermore (γ,1p) reaction cross section is larger about 60 to 886912 than (γ,2p) cross section. Finally, the (γ,2n) cross section is too larger than that of (γ,2p) cross section about 1200 to 10343681 magnitude.

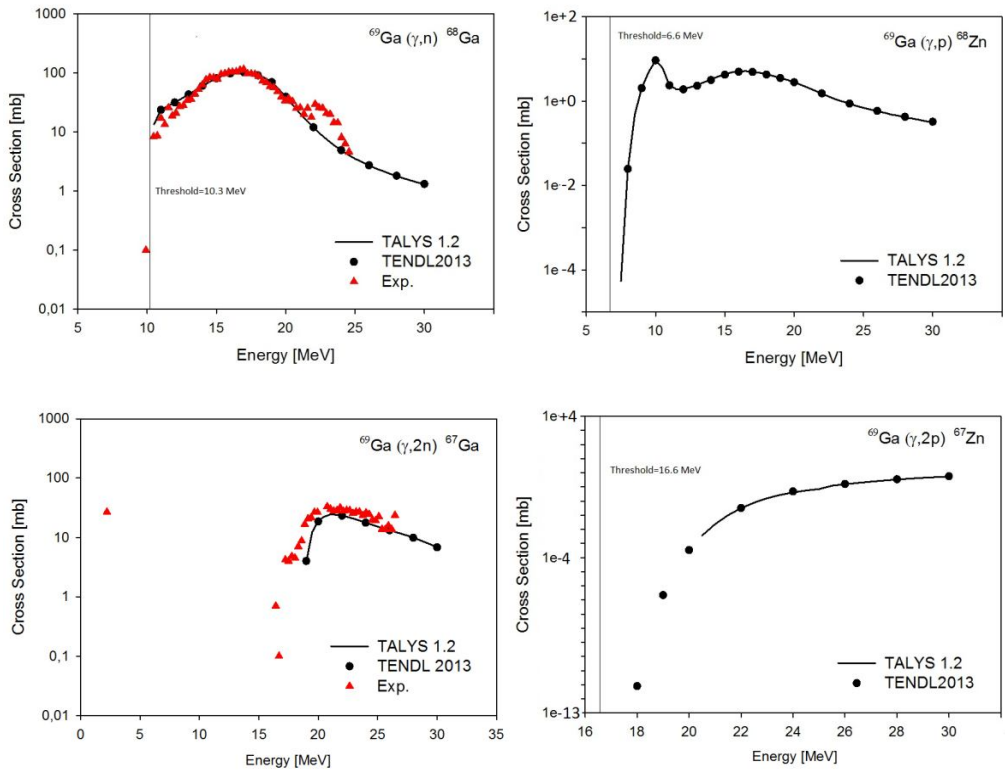

**Fig.2** (γ,1n), (γ,1p), (γ,2n) and (γ,2p) cross sections for $^{69}$Ga nuclei

As can be seen in the Fig.2 that the (γ,1p) reaction cross section sharply increases from threshold energy to about 10.5 MeV where (γ,1n) channel opens up. At this value the increase remains slowly and after about 17 MeV, it becomes decrease because of increasing of the new opening channels such as (γ,2p) and (γ,2n). Besides, the (γ,1n) cross section exhibits similar trend. It increases to 17 MeV and decreases from this point where (γ,2p) and other reaction channels open up.

Similar behavior has been seen for $^{71}$Ga target, but the differences is larger. The (γ,1n) reaction cross section is about 50 mb whilst the (γ,1p) reaction cross section is only about 0.3 mb for the 14 MeV energy value. It is clear in Fig.3 that the photo-neutron cross section is almost 4 to 900 times larger than photo-proton cross section in the energy range 9.5 and 30 MeV. Since there is no explicit difference between neutron and proton thresholds, the reaction probabilities take values in the same energy, 9.5 MeV. As can be seen in the Fig.2 that after 17 MeV, (γ,2n) reaction become taking place. Until 19 MeV the (γ,1n) reaction cross section is larger than that of (γ,2n) reaction. But after this energy value, (γ,2n) become about 1.2 to 7.8 magnitude larger up to 30 MeV. Furthermore (γ,1p) reaction cross section is larger about 1990 to 531000 than (γ,2p) cross section. Finally, the (γ,2n) cross section is too larger than that of (γ,2p) cross section about 71400 to 17400000 magnitude.

Generally, the cross sections for $^{69}$Ga target is larger than that of the cross sections of $^{71}$Ga for (γ,1n), (γ,1p) and (γ,2p) reactions. This difference is slight for (γ,1n) reaction. But for (γ,2n) reaction, the cross section of the reaction performed on $^{71}$Ga target is larger. The cause of this behavior can be interpreted through reaction threshold energies. Because the (γ,2n) reaction threshold energy for $^{71}$Ga is larger than that of the $^{69}$Ga, the probability to happen $^{71}$Ga(γ,2n)$^{69}$Ga reaction is higher. The threshold energies are closer to each other for (γ,1n) reactions, there is a slight difference between the cross sections of the different targets.

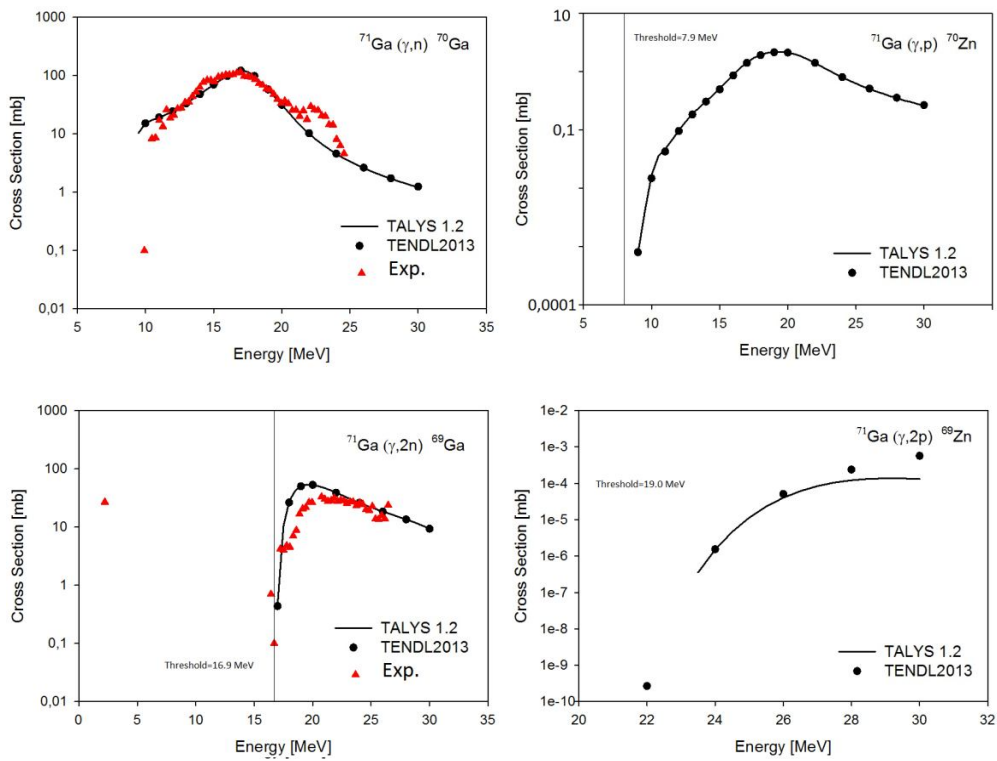

**Fig.3** (γ,1n), (γ,1p), (γ,2n) and (γ,2p) cross sections for $^{71}$Ga nuclei

Since the direct reactions does not contribute to the total cross-section in this energy range for gallium, we have also analyzed the pre-equilibrium and compound mechanism contributions to the total cross section. It is explicit in the Fig.4 that, the compound mechanism is dominant to about 22 MeV for (γ,n) reaction on the $^{69}$Ga target. After this energy value, pre-equilibrium mechanism become dominant to the end. Up to the about 21 MeV, the compound mechanism is again dominant for (γ,p) reaction. For (γ,2n) and (γ,2p) reactions, the compound mechanism is always dominant.

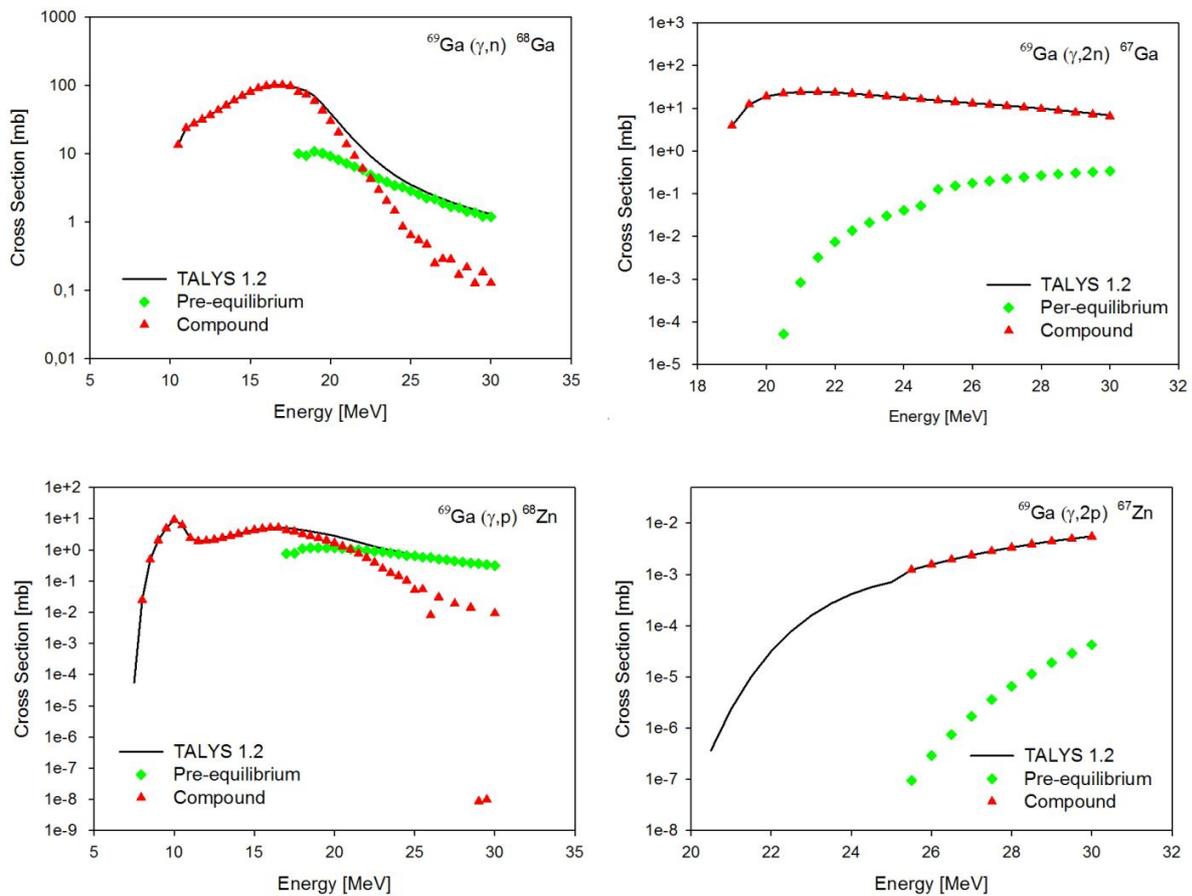

**Fig.4** The contributions of the pre-equilibrium and compound mechanisms to the total cross - sections for (γ,1n), (γ,1p), (γ,2n) and (γ,2p) cross sections for $^{69}$Ga nuclei

Similar analyses have been done for $^{71}$Ga target. It is explicit in the Fig.5 that, the compound mechanism is dominant to about 21 MeV for (γ,n) reaction on the $^{71}$Ga target. After this energy value, pre-equilibrium mechanism become dominant to the end. Up to the about 20.5 MeV, the compound mechanism is again dominant for (γ,p) reaction. For (γ,2n) reaction, the compound mechanism is always dominant. There is no contributions of pre-equilibrium and compound mechanisms to the total cross-section in (γ,2p) reaction.

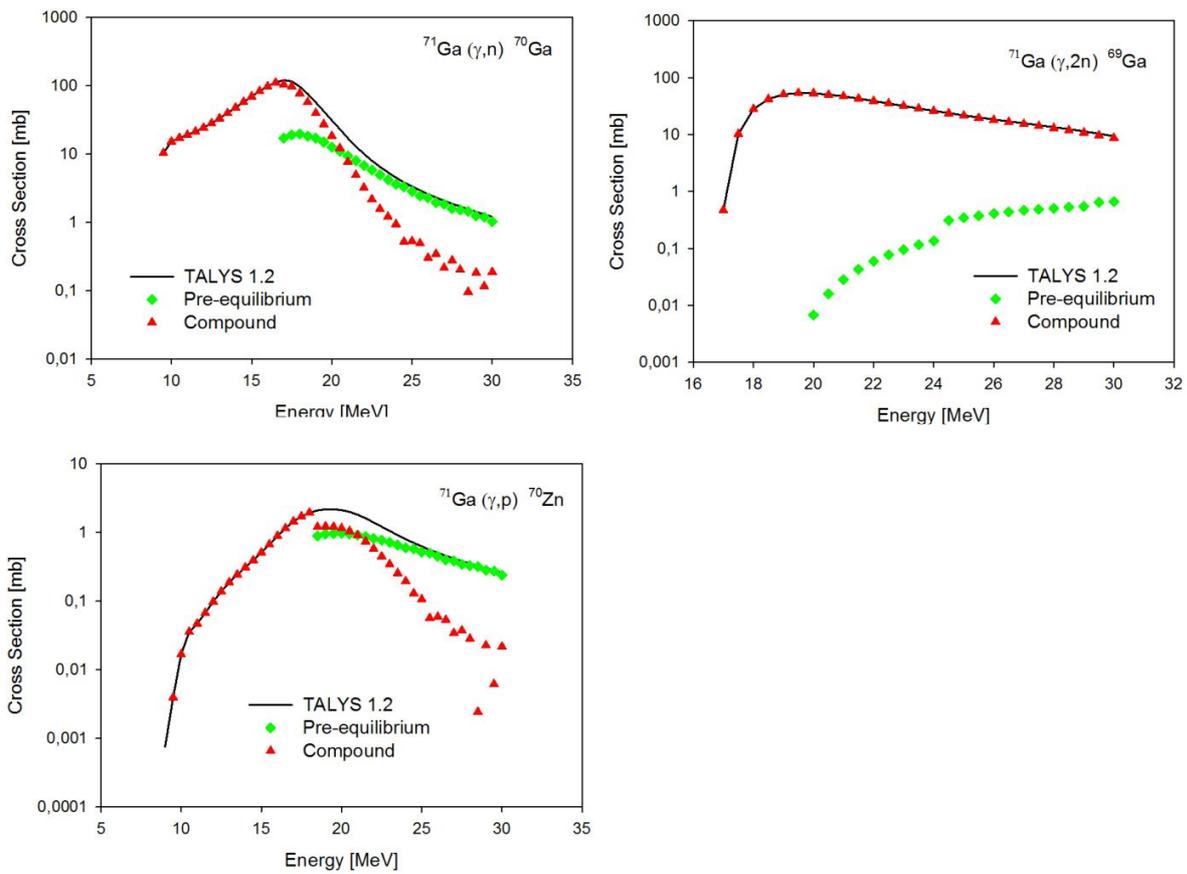

**Fig.5** The contributions of the pre-equilibrium and compound mechanisms to the total cross - sections for (γ,1n), (γ,1p), (γ,2n) and (γ,2p) cross sections for $^{71}$Ga nuclei

**CONCLUSIONS**

The stable gallium isotopes have been considered as a target nuclei in photonuclear reaction simulations. According to the results, we have seen that the calculations are in harmony in the limited literature values. The most probable decay channel is photo-neutron reaction whose cross section is 2 to 20 times larger than that of the photo-proton reaction. By increasing the photon energy, (γ,2n) reaction becomes dominant according to the (γ,n) after about 20 MeV. The pre-equilibrium process is dominant in the high energy region, 20-30 MeV. Besides the compound process contribution to the total cross-section is larger than the others in the 10-20 MeV energy interval.